\documentstyle[sprocl,epsf]{article}

\newcommand{\ewxy}[2]{\setlength{\epsfxsize}{#2}\epsfbox[10 60 640 570]{#1}}

\newcommand{\bee}{\begin{equation}}
\newcommand{\ee}{\end{equation}}
\newcommand{\beea}{\begin{eqnarray}}
\newcommand{\eea}{\end{eqnarray}}

\newcommand{\lsim}{\mathrel{\lower4pt\hbox{$\sim$}}
\hskip-12.5pt\raise1.6pt\hbox{$<$}\;}

\newcommand{\gsim}{\mathrel{\lower4pt\hbox{$\sim$}}
\hskip-12.5pt\raise1.6pt\hbox{$>$}\;}

\begin{document}

\title{ LATTICE QCD AND THE CKM MATRIX}

\author{
Thomas DeGrand}

\address{
Department of Physics\\
 University of Colorado, Boulder CO 80309-390\\ [0.4cm]
}

\maketitle\abstracts{
These lectures provide an introduction to lattice methods for
nonperturbative studies of Quantum Chromodynamics. 
Lecture 1 (Ch. 2) is a very vanilla introduction to lattice QCD.
 Lecture 2 (Ch. 3) describes examples of recent lattice calculations
 relevant to
fixing the parameters of the CKM matrix.
}

\section{Introduction}

The lattice \cite{STANDARD} regularization of QCD
 has been a fruitful source of qualitative and 
quantitative information about QCD for many years, especially when combined
with Monte Carlo simulation.
Lattice methods are presently
the only way we know how to compute masses  and  matrix elements
in the strong interactions beginning with the Lagrangian of QCD.
My goal in these lectures is to give enough of an overview of
the subject that you will be able to make an intelligent
appraisal of a lattice calculation.

The first lecture will describe how to put QCD on a lattice.
 This is a long story with a lot of parts. Lattice QCD is full of
technicalities, but I will try to make the discussion physical.
In Lecture Two I will discuss lattice calculations of matrix elements
which are needed to convert experimental numbers to predictions for the CKM
 matrix.
These calculations have a lot of ingredients: a typical one starts
with a particular choice of discretization and simulation algorithm,
and a choice of operators whose matrix elements are appropriate for
one's measurement. After the lattice number is computed, it may have to
be converted into a number in a continuum regularization scheme (like
$\overline{MS}$), which will involve some kind of perturbative or 
nonperturbative matching calculation. Finally, it might have to be
extrapolated in quark mass, to some physical quark mass value or
 the the chiral limit.  None of these parts are simple or obvious
(or more precisely, most of the time the simple and obvious idea doesn't
 work very well).
Hopefully you will find the physics in the calculations more interesting than
the tables of numbers which result.

\section{Basics of Lattice QCD }
\subsection{Lattice Variables and Actions}

All quantum field theories must be regulated in order to control
their ultraviolet divergences while calculations are performed.
The lattice is a space-time cutoff which eliminates all degrees of freedom
from distances shorter than the lattice spacing $a$. As with any regulator,
it must be removed after renormalization. Contact with experiment only
exists in the continuum limit, when the lattice spacing is taken to zero.
The lattice is a unique regulator compared to the ones you might already know.
 Other regularization schemes are tied closely to perturbative
expansions: one calculates a process to some order in a coupling constant;
divergences are removed order by order in perturbation theory.  The lattice,
however, is a nonperturbative cutoff. Before a calculation begins,
all wavelengths less than a lattice spacing are removed.

All regulators have a price. 
On the lattice we sacrifice all continuous space-time symmetries
 but preserve all internal symmetries, including local gauge 
invariance. 
 This preservation is important for nonperturbative physics. 
 For example, gauge invariance is a property of the continuum theory 
which is nonperturbative, so maintaining it as we pass to the lattice 
means that all of its consequences (including current conservation and 
renormalizability) will be preserved. The bill is paid when we take the
lattice spacing to zero and try to recover what we have left out.

Let's begin by thinking about a lattice version of scalar field theory.
 One just replaces the
space-time coordinate $x_\mu$ by a set of integers $n_\mu$ ($x_\mu=an_\mu$,
where $a$ is the lattice spacing). 
Field variables $\phi(x)$ are defined on sites $\phi(x_n) \equiv \phi_n$,
 The action, an integral over
the Lagrangian, is replaced by a sum over sites
\bee
 S = \int d^4x {\cal L} \rightarrow a^4 \sum_n 
{\cal L}(\phi_n) .\label{2.1}
\ee
and the generating functional for Euclidean Green's functions is
replaced by an ordinary integral over the lattice fields
\bee
Z = \int (\prod_n d \phi_n ) e^{ S}. \label{2.2} 
\ee
Gauge fields are a little more complicated. They carry a
space-time index $\mu$ in addition to an internal symmetry index $a$
($A_\mu^a(x))$ and are associated with a path in space $x_\mu(s)$: a
particle traversing a contour in space picks up a phase factor
\bee
\psi \rightarrow P(\exp \ ig \int_s dx_\mu A_\mu) \psi
 \equiv U(s)\psi(x). \label{2.3}
\ee
$P$ is a path-ordering factor analogous to the time-ordering
operator in ordinary quantum mechanics. Under a gauge transformation $g$,
$U(s)$ is rotated at each end:
\bee
U(s) \rightarrow g^{-1}(x_\mu(s))U(s)g(x_\mu(0)). \label{2.4} 
\ee
These considerations led Wilson \cite{KEN} to formulate gauge fields
on a space-time lattice, in terms of a set of
 fundamental variables  which are elements of the gauge  group $G$ living
on the links of a four-dimensional lattice, connecting neighboring sites
 $x$ and $x+a \mu$:
$U_\mu(x)$, with $U_\mu(x+\mu)^\dagger = U_\mu(x)$
 \bee
U_\mu(n)= \exp (igaT^aA^a_\mu(n))  \label{2.5}
\ee
for $SU(N)$.
($g$ is the coupling,  $A_\mu$ the vector potential, 
and $T^a$ is a group generator).

Under a gauge transformation link variables transform as 
\bee
U_\mu (x) \rightarrow V(x) U_\mu (x) V(x+ \hat \mu)^\dagger  \label{2.10}
\ee
and site variables as 
\bee
\psi(x) \rightarrow V(x) \psi(x) \label{2.11}
\ee
so the only gauge invariant operators we can use as order parameters are 
 matter fields connected by  oriented ``strings" of U's 
\bee
\bar \psi(x_1) U _\mu(x_1)U_\mu(x_1+\hat \mu)\ldots  \psi (x_2)  \label{2.13}
\ee
or closed  oriented loops of U's 
\bee
{\rm Tr} \ldots U _\mu(x)U_\mu(x+\hat \mu)\ldots \rightarrow
{\rm Tr} \ldots U_\mu(x)V^\dagger (x+ \hat \mu)V(x+ \hat \mu)
U_\mu(x+\hat \mu)\ldots  .\label{2.12}
\ee

An action is specified by recalling that the classical Yang-Mills
action involves the curl of $A_\mu$, $F_{\mu\nu}$.
Thus a lattice action ought to involve a product of
$U_\mu$'s around some closed contour. Gauge invariance will
 automatically be satisfied
for actions built of powers of traces of U's around arbitrary closed loops,
with arbitrary coupling constants.
If we assume that the gauge fields are smooth, we can expand the link
variables in a power series in  $gaA_\mu's$. For almost any closed loop, the
leading term in the expansion will be proportional to $F_{\mu\nu}^2$.
This is not a bug, it is a feature. All lattice actions are just bare actions
characterized by many bare parameters (coefficients of loops). In the
continuum (scaling) limit all these actions are in the same universality class,
which is (presumably) the same universality class as QCD with any regularization
scheme, and there will be cutoff-independent predictions from any lattice
actions which are simply predictions of QCD.

Let's hold that thought while we do an example:

 The simplest  contour has a perimeter of four links. In $SU(N)$
\bee
 S={{2} \over {g^2}}\sum_n \sum_{\mu>\nu}{\rm  Re \ Tr \ }
\big( 1 - U_\mu(n)U_\nu(n+\hat\mu)
U^\dagger  _\mu(n+\hat\nu) U^\dagger  _\nu(n) \big).  \label{2.6}
\ee
This action is called the ``plaquette action'' or the
``Wilson action'' after its inventor. $g^2$ is the bare lattice coupling,
whose associated cutoff is $a$.
The lattice parameter $\beta=2N/g^2$ is often written instead of
$g^2=4\pi\alpha_s$.

Let us see how this action reduces to the standard continuum action.
Specializing to the U(1) gauge group, and slightly redefining the coupling,
\bee
S= {1 \over {g^2}}\sum_n \sum_{\mu  > \nu} 
{\rm Re \ }(1 - \exp(iga[A_\mu(n) +A_\nu(n + \hat \mu)
-A_\mu(x+\hat\nu)-A_\nu(n)])).\label{2.7}
\ee
The naive continuum limit is taken by assuming that the lattice spacing 
$a$ is small, and Taylor expanding
\bee
A_\mu(n+\hat\nu) = A_\mu(n) + a \partial_\nu A_\mu(n) + \ldots \label{2.8}
\ee
so the action becomes
\bee
\beta S = {1 \over {g^2}}\sum_n \sum_{\mu > \nu}
1-{\rm Re \ }( \exp(iga[a(\partial_\nu A_\mu -\partial_\mu A_\nu) + O(a^2)])) 
\ee
\bee
={1 \over {4g^2}}a^4\sum_n \sum_{\mu\nu} g^2F_{\mu\nu}^2 + \ldots 
\ee
\bee
= {1 \over 4}\int d^4 x F_{\mu\nu}^2  \\ \nonumber \label{2.9}
\ee
transforming the sum on sites back to an integral.

\subsection{Numerical Simulations}
In a lattice calculation, like any other calculation in quantum field
theory, we compute an expectation value of any observable $\Gamma$
as an average over a ensemble of field configurations:
\bee
\langle{\Gamma}\rangle 
= {1 \over Z} \int [d\phi] \exp(-S)\Gamma(\phi).
\ee
We do this by Monte Carlo simulation: we construct
 an ensemble of states (collection of field variables),
where the probability of finding a particular configuration in the
ensemble is given by Boltzmann weighting (i.~e. proportional to $\exp(-S)$.
Then the expectation value of any observable $\Gamma$ is given simply by
an average over the ensemble:
\bee
\langle{\Gamma}\rangle \simeq \bar \Gamma
\equiv {1 \over N}\sum_{i=1}^N\Gamma[\phi^{(i)}] .
\label{SAMPLE}
\ee
As the number of measurements $N$ becomes large the quantity $\bar \Gamma$
will become a  Gaussian distribution about a mean value, our desired
 expectation value.
The idea  of essentially all simulation algorithms \cite{METRELAX}
is to construct a new configuration
of field variables from an old one.  One begins with some initial field
configuration and monitors observables while the algorithm steps
along. After some number of steps, the value of observables will appear
to become independent of the starting configuration. At that point the
system is said to be ``in equilibrium'' and Eq. \ref{SAMPLE} can be
used to make measurements.

Dynamical fermions are a  complication for QCD \cite{HMDHMC}.
 The fermion path integral is not a number and a computer can't
simulate fermions directly.  However, one can formally integrate out the
fermion fields. For $n_f$ degenerate fermion flavors
\bee
Z = \int [dU][d\psi][d\bar\psi] \exp(-\beta S_G(U) - \sum_{i=1}^{n_f}
\bar \psi M(U) \psi)
\label{ZFERM}
\ee
\bee
=\int [dU](\det M(U))^{n_f}\exp(-\beta S(U)) .
\ee
The determinant introduces a nonlocal interaction among the $U$'s:
\bee
Z = \int [dU] \exp(-\beta S(U)
 - {n_f} {\rm Tr} \ln (M(U))
 ) .  
\ee
Generating configurations of the $U$'s involves computing how the action 
changes when the set of $U$'s are varied. Typically, this involves inverting
the fermion matrix $M(U)$ ($d \log M/dM = M^{-1}$). This is the
 major computational problem dynamical fermion simulations face.
$M$ has eigenvalues with a very large range--
from $2\pi$ down to $m_q a$-- and in the physically interesting limit of
small $m_q$ the matrix becomes ill-conditioned.  At present it is necessary
to compute at unphysically heavy values of the quark mass and to extrapolate
to $m_q=0$. 
(The standard inversion technique today is one of the variants of
the  conjugate gradient algorithm \cite{FROMMER}. ) This tremendous expense
is responsible for one of the``standard'' lattice approximations, the 
``quenched'' approximation. In this approximation the back-reaction of the
fermions on the gauge fields is neglected, by setting
$n_f=0$ in Eq. \ref{ZFERM}.
Valence quarks, or quarks which appear in observables, are kept, but no
sea quarks. No one knows how good an approximation this is, in principle.
In practice it works very well for spectroscopy. The only way we know
how to test it is to compare simulations in the
quenched approximation with those from full QCD.

\subsection{Spectroscopy Calculations}

\begin{figure}
\centerline{\ewxy{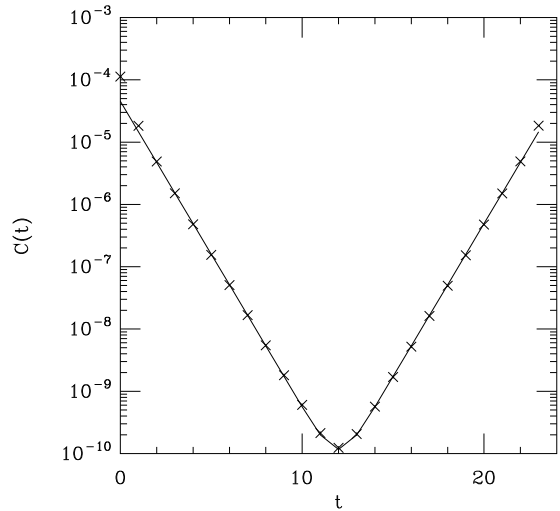}{80mm}
}
\caption{ An obviously very nice looking lattice correlator and its
fit. Periodic boundary conditions convert the exponential decay
into a hyperbolic cosine.}
\label{fig:prop}
\end{figure}

Masses are computed in lattice  simulations from the
asymptotic behavior of Euclidean-time
 correlation functions.  A typical (diagonal) correlator can be written as
\bee
C(t) = \langle 0 | O(t) O(0) | 0\rangle  . 
\label{CT} 
\ee
Making the replacement
\bee
O(t)=e^{Ht}Oe^{-Ht} 
\ee
and inserting a complete set of energy eigenstates,  Eq. \ref{CT} \ becomes
\bee
C(t) = \sum_n |\langle 0 |  O|n\rangle |^2 e^{-E_nt}.  
\ee
At large separation the correlation function is approximately
\bee
C(t) \simeq  |\langle 0 |  O|1\rangle |^2 e^{-E_1t} 
\label{CORRFN}
\ee
where $E_1$ is the energy of the lightest state which the operator $O$
can create from the vacuum. Fig. \ref{fig:prop} shows an example of this.
If the operator does not couple to the vacuum, then
in the limit  of large $t$ one hopes to to find
the mass $E_1$
by measuring the leading exponential falloff of the correlation function.
If the operator $O$ has poor overlap with the lightest state,
a reliable value for the mass can be extracted only at a large time $t$.
In some cases that state is the vacuum itself,
in which $E_1 = 0$.  
Then one looks for the next higher state--a signal which disappears
into the constant background.
This is hard to do.

Most of the observables we are interested in will involve valence fermions.
Let's suppose we wanted to measure the mass of a  meson. Then
we might take 
\bee
C(t) = \sum_x \langle J(x,t) J(0,0) \rangle 
\ee
where
\bee
J(x,t) = \bar \psi(x,t) \Gamma \psi(x,t) 
\ee
and $\Gamma$ is a Dirac matrix.  The intermediate states $|n \rangle$
which saturate $C(x,t)$ are the hadrons which the current $J$ can create
from the vacuum: the pion, for a pseudoscalar
current, the rho, for a vector current,
and so on. 
Now we write out the correlator in terms of fermion fields 
\bee
C(t) = \sum_x \langle 0 | \bar \psi_i(x,t) ^\alpha \Gamma_{ij}
\psi_j(x,t)^\alpha \bar\psi_k(0,0)^\beta 
\Gamma_{kl}\psi_l(0,0)^\beta | 0 \rangle 
\ee
with a Roman index for spin and a Greek index for color.  We contract creation
and annihilation operators into quark propagators 
\bee
\langle 0 | T(\psi_j(x,t)^\alpha \bar \psi_k(0,0)^\beta) | 0 \rangle
 = G_{jk}^{\alpha \beta}(x,t;0,0) 
\ee
so
\bee
C(t) = \sum_x {\rm Tr} G(x,t;0,0) \Gamma G(0,0;x,t) \Gamma 
\ee
where the trace runs over spin and color indices.
Baryons are constructed similarly. A good way to think about these correlators
is by using a sort of Feynman-diagram language which keeps track of
the valence quark lines but ignores all the gluons and sea quarks.

\subsection{The Continuum Limit}

When we define a theory on a lattice the lattice spacing 
 $a$ is an ultraviolet cutoff and all the coupling constants
in the action are  the  bare couplings 
defined with respect it.
  When we take $a$ to zero we must also specify how the couplings
behave.  The proper continuum limit comes when we take $a$ to zero 
holding physical quantities fixed, not when we take $a$ to zero fixing  
the couplings.
 
On the lattice, if all quark masses are set to zero,
 the only dimensionful parameter is the lattice spacing, 
so all masses scale like $1/a$. Said differently, a lattice calculation
produces
the dimensionless combination $am(a)$. One can determine the lattice spacing by
fixing one mass from experiment. Then all other dimensionful quantities
can be predicted.
Imagine computing some masses at several values of the lattice spacing.
(Pick several values of the bare parameters  and
calculate masses for each set of couplings.)
Our calculated mass ratios will depend on the lattice cutoff.
If the lattice spacing is small enough,
 the typical behavior will look like
\bee
(a m_1 (a))/(a m_2 (a)) = m_1(0)/m_2(0) + O(m_1a) + O((m_1 a)^2) +\dots \label{SCALING}
\ee
The leading term does not depend on the value of the UV cutoff, while the
 other terms  do.
The goal of a lattice calculation 
 is to discover the value of some physical observable
as the UV cutoff is taken to be very large, so the physics is in the first
 term.
Everything else is an artifact of the calculation.
We say that a calculation ``scales'' if the $a-$dependent terms in
Eq. \ref{SCALING} are zero or small enough that one can extrapolate
to $a=0$, and generically refer to all the $a-$dependent terms
as ``scale violations.''

We can imagine expressing  each dimensionless combination $am(a)$
as some function of the bare coupling(s) $\{g(a)\}$, $am = f(\{g(a)\})$.
 As $a\rightarrow 0$ we must tune the set of couplings $\{g(a)\}$ so
\bee
 \lim_{a \rightarrow 0}
{1 \over a} f(\{g(a)\}) \rightarrow {\rm constant} . \label{2.35}
\ee
From the point of view of the lattice theory,
 we must tune $\{g\}$ so that correlation lengths $1/ma$ diverge.  
This will occur only at the locations of second (or higher) order phase 
transitions.
In QCD  the fixed point is $g_c = 0 $
so we must tune the coupling to vanish as $a$ goes to zero.

One  needs to set the scale by
taking  one experimental number as input.
A complication that you may not have thought of
 is that the theory we simulate on the computer
is different from the real world. For example, 
the quenched approximation, or for that matter QCD with two
flavors of degenerate quarks,
 almost certainly does not have the same spectrum as QCD
with six flavors of dynamical quarks with their appropriate masses.
Using one mass to set the scale from one of these
approximations to the real world might not give a prediction
for another mass which agrees with experiment.

(The glass is always half empty...In the strong coupling limit, lattice
regularized QCD automatically confines \cite{KEN} and chiral symmetry is
spontaneously broken \cite{CHSB}. So unless there is some kind of phase
 transition as the bare couplings are tuned to take the cutoff away,
which probably doesn't happen, we are working with a confining theory
without doing anything special.)

 Today's 
QCD simulations range from $16^3 \times 32$ to
 $32^3 \times 100$
points and run from hundreds (quenched) to thousands (full QCD) of
hours on the fastest supercomputers in the world. 
 The cost of a Monte Carlo simulation
in a box of physical size $L$ with lattice spacing $a$ and quark mass
$m_q$ scales roughly as
\bee
({L \over a})^4 ({1\over a})^{1-2}({1 \over m_q})^{2-3}
\label{COST}
\ee
where the 4 is just the number of sites, the 1-2 is the cost of
``critical slowing down''--the extent to which successive configurations
are correlated, and the 2-3 is the cost of inverting the fermion
propagator, plus critical slowing down from the nearly massless
pions. Thus it is worthwhile to think about how to do the discretization,
to maximize the value of the lattice spacing.  The thing to keep in mind
is that the lattice action is just a bare action defined with a cutoff.
No lattice discretization is any better or worse (in principle) than any other.
Any bare action which is in the same universality class as QCD will
produce universal numbers in the scaling limit. However, by clever engineering,
it might be possible to devise actions whose scaling behavior is better,
and which can be used at bigger lattice spacing.

\begin{figure}[h!tb]
\centerline{\ewxy{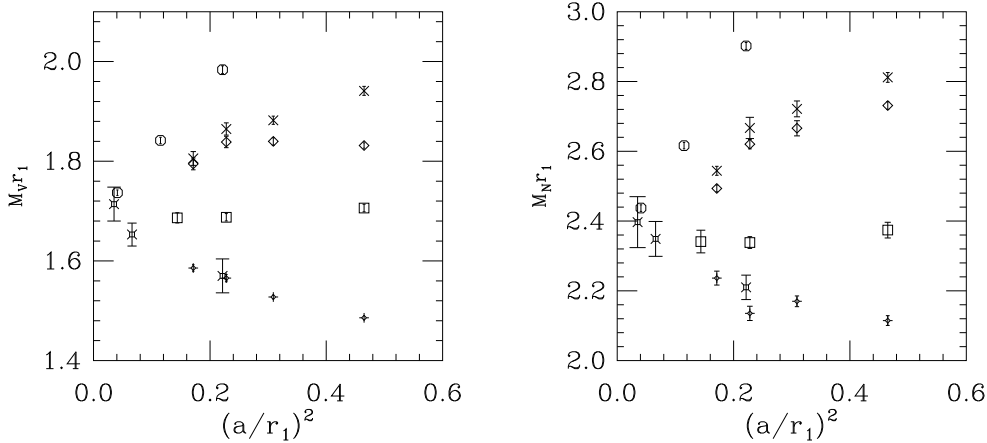}{80mm}
}
\caption{ Lattice calculations of the (a) rho and (b) nucleon mass,
interpolated to the point $m_\pi r_1=0.778$,
as a function of lattice spacing.}
\label{fig:scaling}
\end{figure}

An example \cite{MILCSC}
 of a test of scale violations is shown in Fig. \ref{fig:scaling}.
The x axis is the lattice spacing, in units of a quantity $r_1$,
which is defined through the heavy quark potential:
 $r_1^2 dV(r)/dr|_{r_1}=1.0$, about 0.4 fm. The plotting symbols are for
different kinds of discretizations. The flatter the curve, the smaller
the scale violations.

The simplest organizing principle for ``improvement'' is to
use the  canonical dimensionality of operators as a guide.
Consider the gauge action as an example.
If we perform a naive Taylor expansion of a lattice operator like
the plaquette, we find that it can be written as
\beea
1 - {1 \over 3} {\rm Re \ Tr} U_{plaq} = &
                  r_0 {\rm Tr} F_{\mu\nu}^2  
+a^2 [ r_1 \sum_{\mu\nu}{\rm Tr} D_\mu F_{\mu\nu} D_\mu F_{\mu\nu} +
\nonumber \\
 &  r_2 \sum_{\mu\nu\sigma}{\rm Tr} D_\mu 
F_{\nu\sigma} D_\mu F_{\nu\sigma} + \nonumber \\
 &  r_3 \sum_{\mu\nu\sigma}{\rm Tr} D_\mu F_{\mu\sigma} D_\nu F_{\nu\sigma}]+
 \nonumber \\
 & +O(a^4) 
\eea
The expansion coefficients have a power series expansion in the
coupling, $r_j = A_j + g^2 B_j + \dots$
and the expectation value of any operator $T$ 
computed using the plaquette action will have an expansion
\bee
\langle T(a) \rangle = \langle T(0) \rangle + O(a) + O(g^2 a) + \dots
\ee
Other loops have a similar expansion, with different coefficients.
Now the idea is to take the lattice action to be
a minimal subset of loops and systematically remove the
 $a^n$ terms for physical observables order by order in $n$
 by taking the right linear combination of
loops in the action, $S = \sum_j c_j O_j$
with 
$ c_j = c_j^0 + g^2 c_j^1 + \dots$.
This method was developed by Symanzik and 
co-workers  \cite{SYMANZIK,WEISZ,LWPURE} in the mid-80's.

Ordinary perturbation theory (expansions in the bare lattice coupling $g$)
are not very convergent, but clever prescriptions for definitions of couplings
 \cite{PETERIMP}  or nonperturbative tuning methods\cite{NPIMP}
have been quite successful in developing improved
 lattice actions.

\subsection{Relativistic Fermions on the Lattice}

Defining fermions on the lattice involves yet another problem: doubling.
 Let's illustrate this with free field theory.  The continuum free action is
\bee
S = \int d^4 x [ \bar \psi(x) \gamma_\mu \partial_\mu \psi(x) + m \bar \psi(x)
\psi(x)  ] .\label{2.23}
\ee
One obtains the so-called naive lattice formulation by replacing the
 derivatives
by symmetric differences:
\bee
S_L^{naive} = \sum_{n,\mu} \bar \psi_n {\gamma_\mu \over {2a}}
(\psi_{n+\mu} - \psi_{n-\mu}) + m \sum_n \bar \psi_n \psi_n . \label{2.24}
\ee
The propagator is:
\bee
G(p) = (i \gamma_\mu \sin p_\mu  a + ma)^{-1} 
= {{-i \gamma_\mu \sin p_\mu a + ma}\over{\sum_\mu \sin^2 p_\mu a + m^2 a^2}} 
\label{2.25}
\ee
We identify  the physical spectrum through the poles in the
propagator, at $p_0=iE$:
\bee
\sinh^2 Ea = \sum_j \sin^2 p_j a + m^2a^2
\ee
The lowest energy solutions are the expected ones
 at $p= (0,0,0)$, $E \simeq \pm m$, but  there are 
other degenerate
ones, at $p = (\pi,0,0)$, $(0,\pi,0,)$, \dots $(\pi,\pi,\pi)$. As $a$ goes
to zero, the lightest excitations of the spectrum, the ones whose energy is
 $O(1)$, not $O(1/a)$, are the relevant ones, and there are sixteen of these,
in all the corners of the Bruilloin zone.
Thus our action is a model for sixteen light fermions, not one.
This is the famous ``doubling problem.''

In fact, associated with the ``doubling problem'' is the
Nielsen-Ninomaya \cite{NNTHEOREM} theorem, which says that no lattice
action can be undoubled, chiral, and have couplings which extend over
a finite number of lattice spacings (ultralocality). However, there are
three ways to get two out of three. They are

\noindent
(a) Wilson Fermions (undoubled, nonchiral, ultralocal)

We can alter
 the dispersion relation so that it has only one low energy solution.  The
other solutions are forced to $E \simeq 1/a$ and become very heavy as $a$
is taken to zero.  The simplest version of this solution, called a
 Wilson fermion, adds  an irrelevant operator, a second-derivative-like term
\bee
S^W = -{r \over {2a}}\sum_{n,\mu}\bar \psi_n(\psi_{n+\mu} -2 \psi_n
+\psi_{n-\mu} ) \simeq a r \bar \psi D^2 \psi \label{2.26}
\ee
to $S^{naive}$.  The parameter $r=1$ is 
almost always used and  is implied when one
speaks of using ``Wilson fermions.''

\begin{figure}
\centerline{\ewxy{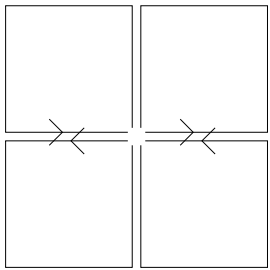}{80mm}
}
\caption{ The ``clover term''.}
\label{fig:clover}
\end{figure}

There are two dimension-five operators which can be added to a fermion
action. The Wilson term is just one of them. The other dimension-five
term is a magnetic moment term
\bee
S_{SW} - {{iag}\over 4} \bar \psi (x)\sigma_{\mu\nu}F_{\mu\nu} \psi(x)
\ee
and if both terms are included, their
 coefficients can be tuned so that there are no
 $O(a g^2)$ lattice artifacts.
This action is called the``Sheikholeslami-Wohlert''  \cite{SHWO} or
  ``clover'' action because the lattice version
of $F_{\mu\nu}$ is the sum of paths shown in Fig. \ref{fig:clover}.

Wilson-type fermions contain an explicit chiral-symmetry breaking term.
This causes a lot of bad things to happen. The most obvious is that
 the zero bare
 quark mass limit is not respected by interactions;
the quark mass is additively renormalized.    The value of 
bare quark mass $m_q$  which the pion mass vanishes, is
not known a priori before beginning a simulation; it must be computed.
This is done in a simulation involving Wilson fermions 
 by varying $m_q$ and watching  the pion mass 
extrapolate quadratically to zero as
$m_\pi^2  \simeq m_q - m_q^c$. It actually turns out that this
is a worse problem than you would think: the Dirac operator $D$ on a
 gauge configuration
could develop a real eigenmode $\lambda$ at minus the bare quark mass you 
dialed into the program. Then $D + m$ would be non-invertible!
Other nasty things happen (operator mixing,
see the next section) and  people
argue about how serious they are in practice \cite{KAPPA}.

\noindent
(b) Staggered or Kogut-Susskind Fermions (chiral, doubled, ultralocal)

\begin{figure}
\centerline{\ewxy{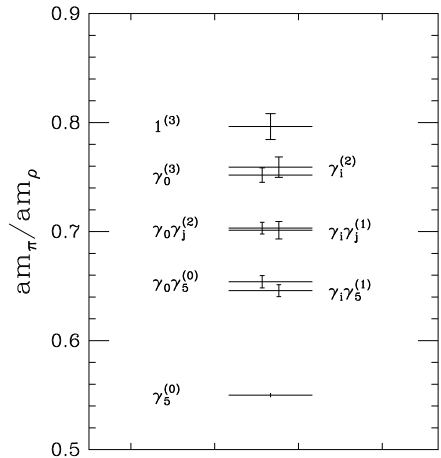}{80mm}
}
\caption{An example of flavor symmetry breaking in an improved
staggered action. The different $\gamma$'s are a code for the various
pseudoscalar states.
Data are from Ref. \protect\cite{OST}. For an explanation of
the splitting, see Ref. \protect\cite{SHARPLEE}.}
\label{fig:flav}
\end{figure}

In this formulation one reduces the number of fermion flavors by using
one component ``staggered'' fermion fields rather than four component Dirac
spinors.  The Dirac spinors are constructed by combining staggered fields
on different lattice sites.
Staggered fermions preserve an explicit chiral symmetry as $m_q \rightarrow 0$
even for finite lattice spacing, as long as all four flavors are degenerate,
although it is not the $SU(N_f)\times SU(N_f)$ of the continuum, it is
a $U(1)$. Thus there is only one Goldstone pion at finite $a$, plus
other non-degenerate pseudoscalar states whose mass goes to zero in
the continuum limit (See Fig. \ref{fig:flav} for an example of this.)
They are preferred over Wilson fermions in situations in which
the chiral properties of the fermions dominate the dynamics.
  They also cheaper to simulate  than Wilson fermions,
because there are less variables.
  However, flavor  and translational
symmetry are all mixed together \cite{GANDSMIT}.

\noindent
(c) Chiral, undoubled, but not ultralocal

These actions implement a modified version \cite{LUSCHER}
 of the chiral rotation
\bee
\delta \psi = \gamma_5 (1 - {1\over 2} a D)\psi; \  \  \
\delta \bar \psi = \bar\psi (1 - {1\over 2} a D)\gamma_5
\ee
which is sufficient to preserve all the interesting features
of continuum chiral symmetry.
An example of such an action is the ``domain wall fermion \cite{BLUM}.'' It is
a variation on the idea that if you have a fermion coupled to a
scalar field, and the scalar field interpolates between two minima
(forms a soliton), the fermion will develop a zero-energy chiral mode
bound to the center of the soliton. Now we go into brane world,
extend QCD into five dimensions, and put ourselves and our four dimensional
world on the kink. There is an anti-kink out there in the fifth dimension,
 and as long as
it is far away the mode on the kink doesn't see the anti-kink and the
4-d
theory on the kink
 is chiral. But if the anti-kink is too close (fifth dimension too small)
the modes mix and chiral symmetry is broken. How close is
 ``too close'' is (yet) another engineering
question.
There are four dimensional analogs of this--think of integrating out
the modes in the fifth dimension in favor of a tower of massive fermions,
and get ``overlap fermions \cite{OVERLAP},'' or construct a low
energy Wilsonian effective action from an underlying chiral theory and get
 ``fixed point fermions \cite{FP}.''
The bad feature is that these actions
have couplings which reach out to many neighboring sites. Their
strength drops exponentially with distance, so they are true local
 actions in the continuum limit, but they are very 
expensive to simulate. But stay tuned...

\section{Hadronic Matrix Elements  from the Lattice }
One of the major goals of lattice calculations is to provide hadronic
matrix elements which either test QCD or can be used as inputs to
test the standard model. 
\subsection{Generic Matrix Element Calculations}
Most of the matrix elements measured on the lattice are
extracted from expectation values of
local operators $J(x)$ composed of quark and gluon fields.
 For example, if one wanted
$\langle 0 | J(x) | h\rangle$ one could look at the two-point function
\bee
C_{JO}(t)= \sum_x \langle 0 | J(x,t) O(0,0) | 0 \rangle .
\label{CURR2}
\ee
Inserting a complete set of correctly normalized momentum eigenstates
\bee
1 = {1 \over L^3} \sum_{A, \vec p}{{|A, \vec p \rangle \langle A, \vec p|}
\over {2E_A(p)}} 
\ee
and using translational invariance and going to large $t$ gives
\bee
C_{JO}(t) = e^{-m_A t} {{\langle 0|J|A\rangle \langle A|O|0\rangle}\over
{2m_A}}. 
\ee
A second calculation of
\bee
C_{OO}(t)= \sum_x \langle 0 | O(x,t) O(0,0) | 0 \rangle
\rightarrow  e^{-m_A t} {{|\langle 0|O|A|\rangle|^2}\over
{2m_A}}
\ee
is needed to extract $\langle 0|J|A\rangle$ by fitting  two
correlators with three parameters.
 
Similarly, a matrix element $\langle h | J | h' \rangle$ can be gotten
from
\bee
C_{AB}(t,t') = \sum_x \langle 0 | O_A(t) J(x,t') O_B(0) | 0 \rangle.
\label{PROTME}
\ee
by stretching the source and sink operators $O_A$ and $O_B$ far apart
on the lattice, letting the lattice project out the lightest states,
and then measuring and dividing out $\langle 0 | O_A |h\rangle$
and $\langle 0 | O_B|h\rangle $.
 
These lattice matrix elements are not yet the continuum matrix elements.
Typically, one is interested in some matrix element defined with
a particular regularization scheme. It is
 a generic feature of quantum field theory that an operator
defined in one scheme ($\overline{MS}$) will be a superposition of
operators in another scheme (lattice). In principle, the superposition
could be all possible operators. So generically an operator of dimension $D$
will mix like
\bee
\langle f | O^{cont}_n(\mu) | i \rangle_{\overline{MS}} =
  a^D\sum_m Z_{nm}\langle f | O^{latt}(a)_m | i \rangle
\label{ZFACTOR}
\ee
 The only restriction are symmetries:
in a theory where parity is conserved a vector operator and an axial vector
operator can't mix. This is relevant for lattice calculations because
the symmetries of the lattice action are in general different from continuum
symmetries.  For example, the space-time symmetry of the lattice
 is given by the group of discrete
 rotations. A more serious source of mixing for light quark 
operators is the way lattice fermions treat chiral symmetry. 
Wilson-type fermions break chiral symmetry (even massless ones do so off-shell)
and so nothing prevents mixing into  ``wrong chirality'' operators.

In Eq. \ref{ZFACTOR} the ``diagonal'' term will contain
the anomalous dimension of the continuum operator
\bee
Z_{nn} = 1 + {g^2 \over{16\pi^2}}(\gamma_n \log \ a\mu + A) +\dots
\ee
(which cancels the mu-dependence of the coefficient function--
$C(\mu)\langle f | O^{cont}|i\rangle_\mu$ is independent of the
 renormalization point). In principle the leading log could be summed,
but in practice we don't know how much of the constant term $A$ should be 
absorbed into a change of scale of $g$, so they are just left there.
 The mixing terms to other dimension $D$ operators
die out in the continuum so they don't have any logs.
There are also terms for mixing with higher dimensional operators, which give
contributions proportional to positive powers of $a$. (These are usually
benign.)
One can also have mixing with lower dimensional operators, with contributions
involving negative powers of $a$. (Four fermion operators for $B_K$
could mix with $s\bar d$.) These are deadly. They must drop out in the
continuum but it is a delicate business, since they look like they are
growing as an inverse power of $a$.

This is probably more than you wanted to know, but you
 need the $Z_{nm}$'s to produce numbers. People get them in a number
 of ways. Most straightforward is to compute them in perturbation theory,
but lattice perturbation theory in terms of the bare coupling $g(a)$
is not very convergent, and it is a long tricky story to do better.
The culprit is the ``tadpole graph.'' 
The lattice fermion-gauge field interaction is generically
$\bar \psi(x) U_\mu(x) \psi(x+\hat mu)$ and $U \simeq 1 +iga A_\mu - g^2a^2/2
A_\mu^2 + \dots$. The $\bar \psi A_\mu^2 \psi$ vertex, not present in
any sensible continuum regularization, causes problems when the gluon
forms a loop: the quadratic divergence from the loop integral
 combines with the $a^2$ to give
a finite contribution--in fact, it is often the dominant contribution.
In perturbation theory one must also choose the momentum scale in the
running coupling constant. There are reasonable choices for how to do 
that \cite{BLM,PETERPAUL}.

Often one can find $Z_{nm}$'s by forcing  lattice observables to obey Ward
identities \cite{MM}.
One can also play this game with quark propagators and vertices,
 by computing analogs of quark vertices on the lattice
and matching ones results to a continuum calculation \cite{MART}.

Besides, the $Z$'s, there are
other things that can go wrong.  Most  lattice actions
break down when the quark mass gets heavy. The dispersion relation for
Wilson or clover actions is $E(p) = m_1 + p^2/(2m_2)$ and the quark magnetic
moment is $\mu=1/m_3$ with $m_1 \ne m_2 \ne m_3$.
The residue of the quark propagator at its pole is not $1/(2E)$ as in the
continuum. What to do then is not  obvious (meaning that lattice
people fight over what to do).

\subsection{Heavy quark operators}
There are many lattice calculations of $f_B$, $f_D$, $B_B$, and form factors
for semileptonic decay.
 $\bar B-B$ mixing is parameterized by the
ratio $x_d  = {{(\Delta M)_{b \bar d}}/ {\Gamma_{b \bar d}}}$
\bee
x_d  = 
\tau_{b \bar d}{{G_F^2}\over{6\pi^2}}\eta_{QCD}F\big({{m_t^2}\over{m_W^2}}\big)
 |V_{tb}^*V_{td}|^2 
 b(\mu) \{ {3\over 8}\langle \bar B| \bar b \gamma_\rho(1-\gamma_5) d
\bar b \gamma_\rho(1-\gamma_5) d | B \rangle  \}
\label{BIGEQ}
\ee
Experiment is on the left; theory on the right. Moving into the
long equation from the left, we see many known (more or less) parameters
from phase space integrals or perturbative QCD calculations, then
a combination of CKM matrix elements, followed by a four quark hadronic
matrix element \cite{ROSNER}. We would like to extract the CKM
matrix element from the measurement of $x_d$ (and its strange partner
$x_s$). To do so we need to know the value of the
 object in the curly brackets, defined as $3/8 M_{bd}$
and  parameterized as $m_B^2 f_{B_d}^2 B_{b_d}$ where $B_{b_d}$ is the
so-called B-parameter, and $f_B$ is the B-meson decay constant,
$\langle 0 | \bar b \gamma_0 \gamma_5 d | B \rangle =f_Bm_B$.
Vacuum saturation suggests that
$B_B=1$. From the lattice one can try to get a real value.

In Eq. \ref{BIGEQ}  $b(\mu)$, the coefficient which runs the
effective interaction down from the W-boson scale to the QCD scale $\mu$,
and the matrix element $M(\mu)$ both depend on the QCD scale.
One often sees the renormalization group invariant quantities
$\hat M_{bd} = b(\mu) M_{bd}(\mu)$ or $\hat B_{bd} = b(\mu) B_{bd}(\mu)$
quoted in the literature.

Decay constants  probe very simple properties of the wave function: in the
nonrelativistic quark model
$
f_M = {{\psi(0)}/{\sqrt{m_M}}} 
$,
where $\psi(0)$ is the $\bar q q$ wave function at the origin.
For a heavy quark ($Q$) light quark ($q$) system $\psi(0)$ should become
independent of the heavy quark's mass as the $Q$ mass goes to infinity, and
in that limit one can show in QCD that $\sqrt{m_M}f_M$ approaches a constant.

The decay constant is computed by combining a heavy quark and a light antiquark
propagator into Eq. \ref{CURR2}. You might think it would be
 difficult to calculate $f_B$ directly on present day lattices
with relativistic lattice fermions
because the lattice spacing is much greater than the $b$ quark's
Compton wavelength (or the UV cutoff is below $m_b$).  But it is better
to think of the lattice theory as an effective field theory for 
the low-momentum excitations in the presence of additional high
energy scales--the cutoff (inverse lattice spacing) and the heavy quark mass.
As in any effective field theory, the effects of the short distance
are lumped into coefficients of the effective theory
\cite{FERMILAB}.
As a practical matter, one can use the good old clover action to do the
calculations--it contains all the necessary operators.
 The bare mass has nothing direct to do with the results; one
 tunes it, monitoring the kinetic energy $E(p)= m_1 + p^2/(2m_2)+\dots$,
and takes the hadron mass to be $m_2$.

Nonrelativistic QCD has also been discretized and used to make very
 precise calculations of the properties of quarkonia \cite{REF6}.
 This formalism can
also be used for the heavy quark (again as long as its momentum is small.)
The ``static'' limit (infinite $b$-quark mass)
 is often used as an additional point on the
curve.  
 Then one can  try to extrapolate all the way from light quarks to heavies
and get all the decay constants at once. 
 
I will show some pictures from the lattice decay constant of Ref.
 \cite{BERNARD}.
These authors (my name is on it but I didn't do anything)
 did careful quenched simulations at many values of the lattice
spacing, which allows one to extrapolate to the continuum limit by brute
force. They have also done a  set of simulations which
include light dynamical quarks, which should give some idea of the
accuracy of the quenched approximation.

\begin{figure}[thb]
\epsfxsize=0.8 \hsize
\epsffile{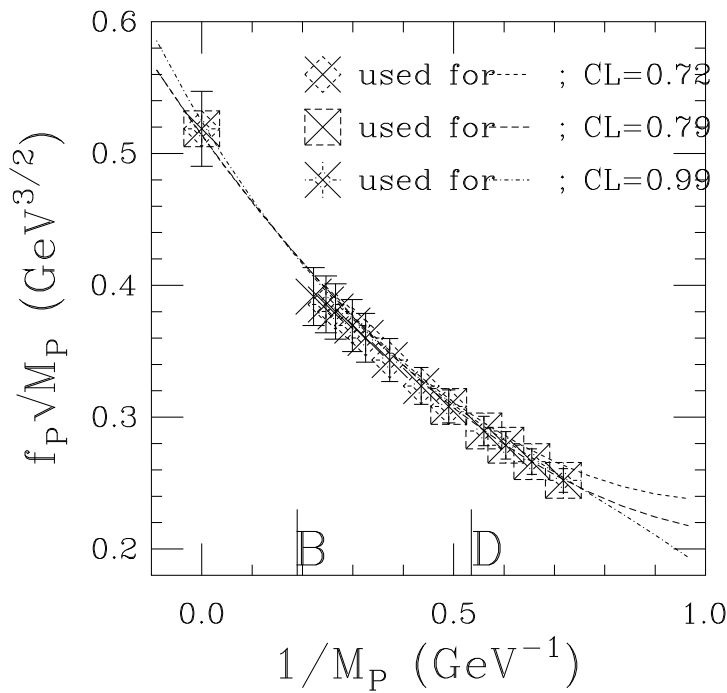}
\caption{
Pseudoscalar meson decay constant vs $1/M$, from Ref.\protect\cite{BERNARD}.}
\label{fig:frootm}
\end{figure}

\begin{figure}[thb]
\epsfxsize=0.8 \hsize
\epsffile{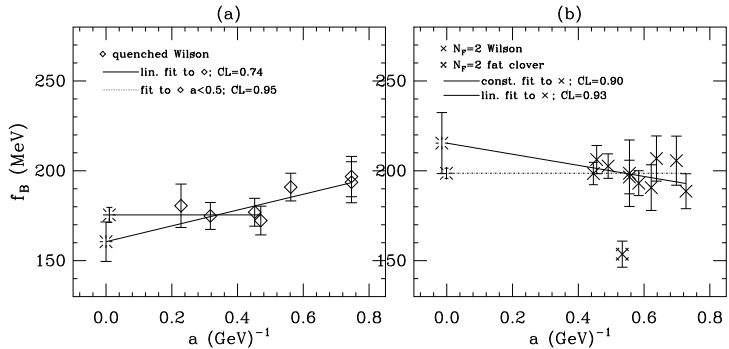}
\caption{ $f_B $ vs. $a$ from Ref. \protect\cite{BERNARD}, 
showing extrapolations to
the continuum limit of quenched (a) and full (b) QCD data.
The scale
is set by $f_\pi=132$ MeV throughout.
}
\label{fig:fb}
\end{figure}

Examples of  results of Ref. \cite{BERNARD} are shown in Figs.
\ref{fig:frootm} and \ref{fig:fb}.
The simulations with dynamical fermions are not as good quality
as the quenched simulations: the lattice spacings are generally larger,
the simulations all have two degenerate flavors (what about the strange quark),
and the dynamical  quark masses are still a bit large. We think that
the Wilson results (crosses in Fig. \ref{fig:fb}(b) over estimate the
continuum result,
and the clover action we are using underestimates it, but
we also suspect  that quenched $f_B$ is a bit too low.

Soni \cite{SONI} has presented a summary of data from various collaborations,
as of last winter. Again, there is a hint that the $N_f=2$ results may be 
about 30 MeV above the quenched results.

\begin{table}[tbh]
\caption{Heavy-light decay constants and their
ratios. \label{tabone}}
\begin{center}
\begin{tabular}{|l|c|c|}
\hline
Quantity & Quenched ($N_f=0$) & Partially Unquenched ($N_f=2$) \\
\hline
$f_B$/MeV & $170\pm20$ & $200\pm30$ \\
$f_{B_S}$/MeV & $190\pm20$ & $220\pm30$ \\
$f_D$/MeV & $205\pm20$ & $225\pm30$ \\
$f_{D_S}$/MeV & $225\pm20$ & $245\pm30$ \\
$f_{B_S}/f_B$ & $1.14\pm.06$ & $1.14\pm.06$ \\
$f_{D_S}/f_D$ & $1.10\pm.06$ & $1.10\pm.06$ \\
\hline
\end{tabular}
\end{center}
\end{table}

Lattice calculations have been predicting quenched
$f_{D_s} \simeq 200$ MeV for about twelve years.  
The central values have changed very little, while the uncertainties
have decreased. Experimental determinations of $f_{D_s}$ all come in
higher than the lattice results, though with large error bars.
We need to do a good quality unquenched lattice calculation.

\begin{figure}[thb]
\epsfxsize=0.8 \hsize
\epsffile{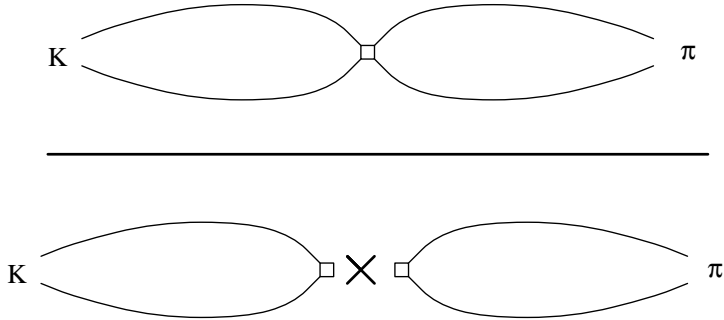}
\caption{ $B_K$ (shown) and $B_B$ are computed by taking the ratio of 
four quark operator and two two-point functions.
(Figure from Ref. \protect\cite{PANDK}.)}
\label{fig:ratio}
\end{figure}

Now back to the $B$ parameter. On the lattice, one could measure 
the decay constants and $B$ parameter separately and combine them after 
extrapolation, or measure $M$ directly.
In principle the numbers should be the same, but in practice the first 
technique has produced better numbers so far. That is because the 
$B$ parameter is measured as the ratio of a correlator with a four-fermion
vertex to a product of two current-current correlators
 (see Fig. \ref{fig:ratio}).
A lot of systematics cancel in the ratio.

Many groups have visited this problem. Reviews by Draper \cite{DRAPER} 
and Soni \cite{SONI} quote a world summary, which I copy into Table
\ref{tabthree}.

\begin{table}[tbh]
\caption{Summary of heavy-light $B$-parameters.
\label{tabthree}}
\begin{center}
\begin{tabular}{|l|c|c|}
\hline
 & Quenched &  ``Unquenched''  \\
\hline
$B_{B_d}(m_b)$ & .86(4)(8) & .86(4)(8) \\
$B_{B_S}/B_{B_d}$ & 1.00(1)(2) & 1.00(1)(2) \\
$f_{B_d}(\hat B^{nlo}_{B_d})^{1/2}$ & $195\pm25$ MeV & $230\pm35$ \\
$\frac{f_{B_S}(B^{nlo}_{B_S})^{1/2}}{f_{B_d} (B^{nlo}_{B_d})^{1/2}}$ &
$1.14\pm.06$ & $1.14\pm.07$ \\
\hline
\end{tabular}
\end{center}
\end{table}

Semileptonic decays involve processes like Eqn. \ref{PROTME}.
On the lattice, one just measures the matrix element of
a current and fits it to the expected set of form factors--for
$B\rightarrow \pi \ell \nu$, for example,
\bee
\langle \pi(p) | V_\mu |B(p')\rangle =
 f^+(q^2)[p'+p - {{m_B^2-m_\pi^2} \over q^2}q]_\mu +
 f^0(q^2){{m_B^2-m_\pi^2} \over q^2}q_\mu
\ee
 The best signals come when the momenta of the initial
and final hadron are small.
 Then the large $B$ mass forces
$q^2$ ($q={}$lepton 4-momentum${}=p_B-p_\pi$) to be  large.
If the form factor is needed  at 
$q^2\sim0$, a large extrapolation is needed,
and there will be additional errors and model dependence in the answer.
(Lattice people have no advantage over anyone else at guessing at
functional forms.)
  However, finding $V_{ub}$
from experimental data only requires knowing the form
factor at one value of $q^2$. This should work so long as the experiment has
enough data to measure the differential rate 
 around that region of $q^2$.
Two recent
approaches try to do this:
UKQCD focussed on near the end-point or the zero-recoil region where
the lattice data tends to be cleanest and heavy quark symmetry can be used.
The FNAL group has measured $B\to D\ell\nu$ form factors at zero 
recoil \cite{FNALPRD}. They have a clever technique from removing much
of the lattice-to-continuum Z-factors by computing ratios of
matrix elements, such as
\bee
{{
\langle D | \bar c \gamma_0 b | \bar B \rangle
\langle \bar B | \bar b \gamma_0 c |  D \rangle
}\over{
\langle D | \bar c \gamma_0 c | D \rangle
\langle \bar B | \bar b \gamma_0 b | \bar B \rangle
}} = |h_+(v\cdot v'=1)|^2
\ee
The denominators are just diagonal matrix elements of the charge density,
and they can easily be normalized.
They \cite{ryan} are also computing
semi-leptonic form factors for $B\to\pi\ell\nu$ and $D\to \pi(K)\ell\nu$,
by concentrating directly on the differential decay
spectrum in an interval with $0.4\lsim \vec p_\pi/{\rm GeV}\lsim0.8$
thus avoiding the need for large extrapolation in $q^2$.

\subsection{Kaon  Matrix Elements}

Lattice calculations of kaon weak interaction matrix elements begin
with the full Standard Model at high energies and use the operator
product expansion, combined with the renormalization group, to construct
a low-energy effective field theory valid at scales $\mu$ of a few GeV.
The effective Hamiltonian basically reduces to a sum of four-fermion
interactions
\bee
H_W^{eff} = {G_F \over {\sqrt{2}}}\sum_{i=0}^{10} c_i(\mu) O_i(\mu)
\ee
People have expended the most effort on, and have the best results for, $B_K$;
there are some results on the $\Delta I = 1/2$ rule; and last,
there is
$\epsilon'/\epsilon$, with unreliable results so far.

\noindent{\underline {$B_K$}}

\begin{figure}[thb]
\epsfxsize=0.8 \hsize
\epsffile{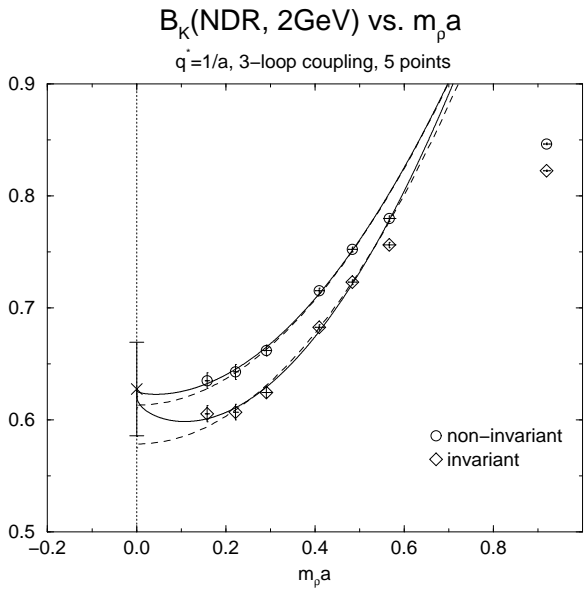}
\caption{ $B_K$ from staggered fermions, as a function of the lattice spacing,
for two different choices of lattice operators.
(Figure and results from Ref. \protect\cite{JLQCDKS}.)}
\label{fig:bkjlqcd}
\end{figure}

The JLQCD collaboration has the best results on $B_K$,
from a calculation using staggered fermions\cite{JLQCDKS}. They have
data from many lattice spacings and several choices for the
 lattice discretization of the operator. (See Fig. \ref{fig:bkjlqcd}.)
They find quenched $B_K(\overline{MS},\mu=2$ GeV) = 0.616(5).
The main limitations of this result are quenching, plus the fact that
the lattice calculations are actually done without $SU(3)$ flavor
breaking (the lattice ``kaon'' is a pseudoscalar made of degenerate
quarks).  These effects are believed \cite{SONI}
 to be 5-10 per cent corrections.

JLQCD \cite{JLQCDW} has  also done a calculation with Wilson  fermions.
This was done not so much to get a number itself but to check the staggered
result. The systematics are very different and the operator mixing is fierce
due to the loss of chiral symmetry inherent in Wilson fermions.
For example,
 $\bar s \gamma_\mu (1-\gamma_5) d \cdot\bar s \gamma_\mu (1-\gamma_5) d$ 
mixes with 
 $\bar s \gamma_5  d \cdot\bar s \gamma_5 d$ and that operator
has a $\bar K-K$ matrix element ten times greater.

\noindent{\underline {$\Delta I = 1/2$ Rule}} 

\begin{figure}[thb]
\epsfxsize=0.8 \hsize
\epsffile{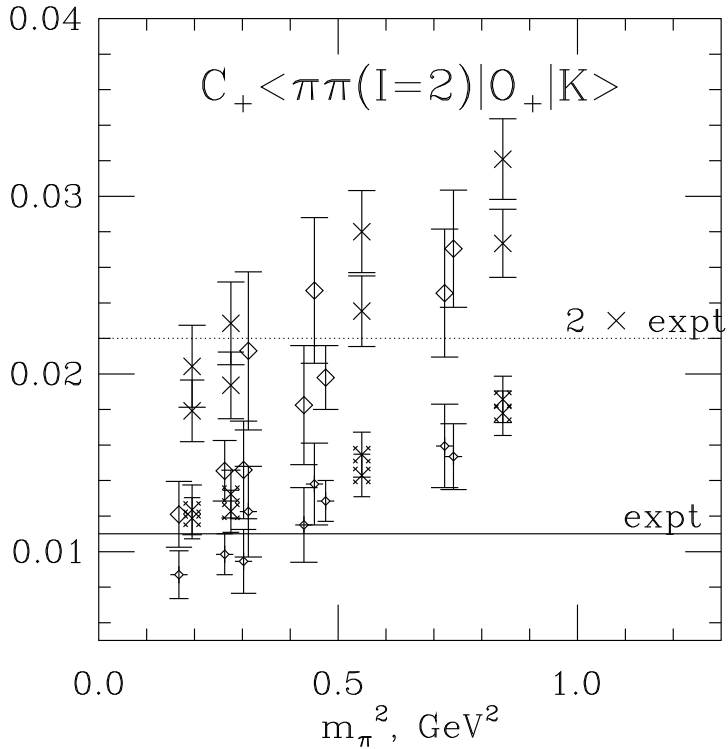}
\caption{$\Delta I = 3/2$ $K\rightarrow \pi\pi$ amplitude with (fancy symbols)
and without (plain symbols) one-loop corrections of quenched chiral
 perturbation theory. Data are
 crosses and fancy crosses, \protect\cite{BERNA};
 diamonds and fancy diamonds, \protect\cite{JLDI}.
Data are plotted as a function of lattice meson mass.
This figure (and much else) is from \protect\cite{KURA}.
}
\label{fig:di32}
\end{figure}

Can the lattice reproduce the experimentally observed factor of 
22 between $K^0 \rightarrow (\pi\pi)_{I=0}$ and 
$K^0 \rightarrow (\pi\pi)_{I=2}$ amplitudes?
The lattice calculations are difficult. In addition to graphs of
 Fig. \ref{fig:ratio},
which are reasonably straightforward to compute, there are a host of other
topologies,
 some of which involve computing propagators from many points on the
lattice to many other points.  But I think the reason there are so few
 lattice results is because all  of the quantities of interest
 are scheme dependent and
one must compute a lattice-to-continuum matching factor.
In addition, people don't calculate $K\rightarrow \pi\pi$ directly on the
 lattice; it is difficult \cite{MAINIETAL} to extract the phase shifts
from the $\pi\pi$ final state interactions from lattice data (never mind
trying to separate the two pions to asymptotically great distances).
Instead, they use chiral perturbation theory \cite{CURRAL} to relate
$K\rightarrow \pi\pi$ amplitudes to $K\rightarrow \pi$. In the case of the
$\Delta I =3/2$ amplitude there is a factor of two change in the lattice
result depending on whether tree level or one loop chiral perturbation theory
is used. This is shown in Fig. \ref{fig:di32}.

\noindent{\underline {$\epsilon'/\epsilon$}}

The only recent work I am aware of is by Pekurovsky and Kilcup \cite{PANDK}.
The calculation is hard.
 The biggest operators $O_6$ and $O_8$ in the nomenclature
have opposite signs and nearly cancel. But the big problem is the
scheme matching. In a perturbative calculation,
we saw that $\langle O \rangle_{\overline{MS}} \simeq Z
 \langle O \rangle_{latt}$ and $Z= 1 + \alpha_s C + \dots$. But in this
equation, what scheme is used to compute $\alpha_s$, and what
is the scale $q^*$ at which $\alpha_s$ is evaluated? Pekurovsky and Kilcup
found that their  numbers for one operator, $O_6$, shifted by a factor of
4 when they were converted into $\overline{MS}$, and the factor of 4 could
become a factor of 30 (or worse) as $q^*$ was varied from $\pi/a$ to $1/a$.
They attempt to guesstimate
 numbers but since they say plainly that  they should be used
``with extreme caution'' I won't quote them.
A nonperturbative approach to matching
 is clearly needed but does not exist yet.

\section{Conclusions}

What about the future?
Matrix elements are at the end of a long chain involving a large set of
 both  simulation
and physics issues. They are the most complicated corner of
the lattice game. If all you want are the numbers,
Moore's law says that computer speed doubles 
every eighteen months, and statistics is $\sqrt N$, so error bars will
fall by a factor of 2 every three years for everything we know how
to do today and will learn nothing new about how to do better in the future.
And there are many projects and proposals to build clusters or
dedicated supercomputers
at a cost which is ``chicken feed'' compared to the experimental program.
This will enable us to begin to chip away at
 the biggest systematic in all the calculations
I have shown here--the neglect of the quenched approximation.

But new hardware is not really where the action is. It is
 merely ``enabling technology,'' so we can make mistakes faster, learn
more about the physics, and test new ideas.

The main bottleneck to progress on hadronic matrix elements is just
that these calculations are complex and have many parts.
Some of us (me, let's not be shy)
 think that better discretization algorithms will help.
The problem with that approach is that many pieces of the puzzle
have to be determined from scratch: learning how to optimize
 the new algorithm, testing spectroscopy, computing the $Z$'s. This
takes a couple of years, if the inventor of the algorithm doesn't get tired
first.
 Others of us prefer to live with poorer algorithms (which have already been
well calibrated) and try to tweak the parts of them which work the worst.
The simulations still take a couple of years.
Believe it or not, even though lattice QCD is a mature field, there
 are still many questions about QCD which lattice people
do not know how to answer, and an outsider might.
 Maybe you would enjoy thinking about them.

\section*{Acknowledgements}
I would like to thank 
S.~Gottlieb,
Y. Kuramashi,
A. Kronfeld,
M. Ogilvie,
S. Sharpe,
A. Soni,
and
D.~Toussaint
for their help preparing these lectures.
This work was supported by the U.~S. Department of Energy.

\newcommand{\PL}[3]{{Phys. Lett.} {\bf #1} {(19#2)} #3}
\newcommand{\PR}[3]{{Phys. Rev.} {\bf #1} {(19#2)}  #3}
\newcommand{\NP}[3]{{Nucl. Phys.} {\bf #1} {(19#2)} #3}
\newcommand{\PRL}[3]{{Phys. Rev. Lett.} {\bf #1} {(19#2)} #3}
\newcommand{\PREPC}[3]{{Phys. Rep.} {\bf #1} {(19#2)}  #3}
\newcommand{\ZPHYS}[3]{{Z. Phys.} {\bf #1} {(19#2)} #3}
\newcommand{\ANN}[3]{{Ann. Phys. (N.Y.)} {\bf #1} {(19#2)} #3}
\newcommand{\HELV}[3]{{Helv. Phys. Acta} {\bf #1} {(19#2)} #3}
\newcommand{\NC}[3]{{Nuovo Cim.} {\bf #1} {(19#2)} #3}
\newcommand{\CMP}[3]{{Comm. Math. Phys.} {\bf #1} {(19#2)} #3}
\newcommand{\REVMP}[3]{{Rev. Mod. Phys.} {\bf #1} {(19#2)} #3}
\newcommand{\ADD}[3]{{\hspace{.1truecm}}{\bf #1} {(19#2)} #3}
\newcommand{\PA}[3] {{Physica} {\bf #1} {(19#2)} #3}
\newcommand{\JE}[3] {{JETP} {\bf #1} {(19#2)} #3}
\newcommand{\FS}[3] {{Nucl. Phys.} {\bf #1}{[FS#2]} {(19#2)} #3}


\begin{thebibliography}{10}
\bibitem{KAPPA}
For technical reasons, lattice people using Wilson or clover fermions
like to work with 
the ``hopping parameter,''  universally labeled $\kappa$. It
is related to the bare quark mass by $\kappa = 1/(2m_q + 8)$.
The pion is massless at $\kappa=\kappa_c$.
\end{thebibliography}
\end{document}